\documentstyle[preprint,eqsecnum,aps]{revtex}
\begin{document}
\draft
\preprint{KAIST--TH 03/98, SNUTP  98-008}
\title{Recent Developments In 
Heavy Quarkonium Physics   }
\author{Pyungwon Ko
\thanks{ pko@chep6.kaist.ac.kr}
\footnote{Talk given at PPPP workshop, Seoul, Korea (Oct. 31 -- Nov. 2, 1997)
}  \\ }
\address{Dep. of Physics, KAIST \\ Taejon 305-701, Korea 
}
\maketitle
\begin{abstract}
Recent developments in heavy quarknoium physics are reviewed in brief, 
including (i) nonrelativistic QCD(NRQCD), (ii) the importance of color-octet 
$(Q\bar{Q})$ components in the decay and the production of a physical heavy 
quarkonium state, (iii) limitation of the NRQCD factorization and (iv) the 
double humps in the $\pi\pi$ spectrum in the decay $\Upsilon(3S) \rightarrow 
\Upsilon (1S) \pi\pi$. 
\end{abstract}

\tighten
\narrowtext

\section{Introduction}

Heavy quarkonium physics was one of the early applications of perturbative 
QCD (PQCD). In the original treatment of the decay and the production of a 
heavy quarkonium state $H$, it was assumed that $(Q\bar{Q})_{1}$ inside $H$ 
is in a definite $^{2S+1}L_J$ color-singlet state. And  production and 
decay rates of a heavy quarkonium $H$ are assumed to be factorized into 
(i) the short distance (SD) parts which are calculable in PQCD (in $\alpha_s 
(M_Q)$), and (ii) the long distance (LD) parts that may be parametrized 
in terms of the $(Q\bar{Q})_1$  wave function and its derivatives at the 
origin \cite{csm}.  

This factorization hypothesis works well for the $S-$wave and $P-$wave states 
to lowest order in $\alpha_s$. For example, the $\chi_{c0}$ state decays 
into light hadrons (LH) through $\chi_{c0} \rightarrow g g$ 
at the parton level : 
\begin{equation}
\Gamma ( \chi_{c0} \rightarrow {\rm LH}  )  = 
{18 \alpha_s^2 (M_c) \over M_c^4 }~| R_{p}^{'} (0) |^2.
\end{equation}
However, such a factorization in the CSM fails if PQCD correction is 
included, which is signaled by the infrared divergence 
in its decay rates \cite{chi0}. 
This infrared divergence implies the failure of the factorization in terms 
of a single wave function of $(Q\bar{Q})_1$ in $H$. This problem remained 
unsolved until 1992, when Bodwin {\it et al.} pointed out {\it the 
importance of the  color-octet} $(c\bar{c})_8$ {\it component} of $H$ 
(see (5)) \cite{bbl92}.   

%

In this decade,  the prompt $J/\psi$ production at the Tevatron could be 
measured with the development of high resolution vertex detector.  The data
was astonishing in that the CSM underestimates the data by a factor of $\sim
10$ for $J/\psi$ production, even worse for  $\psi^{'}$ production (by 
$\sim 30$). Possible contributions in the CSM are (i) gluon-gluon fusion 
($g g \rightarrow J/\psi + g$), (ii)  gluon fragmentation ($g \rightarrow 
J/\psi + g g$), (iii) $c-$quark fragmentation  ($c \rightarrow J/\psi + c$),
and (iv) cascade decays from $P-$wave charmonia ($\chi_{cJ} (1P) \rightarrow 
J/\psi + \gamma$), which is the most important for the $J/\psi$ productions.
However, for the case of  $\psi^{'} $ production at the Tevatron, one 
observed that $\sigma_{\rm exp} \sim 30 \times \sigma_{\rm th}$ based on  
CSM. Since there are no known $P-$wave states (such as $\chi_{cJ} (2P)$) 
that can decay into $\psi^{'}$, two options are avaiable : 
(i) hypothetical $\chi_{cJ}(2P)$ decaying  into $\psi^{'} + \gamma$ with a 
suitable branching ratio \cite{roy}, and can be tested in $B$ and $\Upsilon$
decays \cite{ko0}, and 
(ii) color-octet gluon fragmentation : $g \rightarrow (c\bar{c})_8$ 
followed by $(c\bar{c})_8 \rightarrow J/\psi + {\rm soft}~g g$ \cite{fleming}.
This second idea is the main theme of this talk, and the next two sections
will be devoted to this issue. 

Another less understood phenomenon in the heavy quarkonium physics in 1980's
is the  double humps in $\pi\pi$ spectrum in $\Upsilon(3S) \rightarrow 
\Upsilon(1S) \pi\pi$ \footnote{although this puzzle has nothing to do 
with the limitation of the CSM}. This issue is related with chiral dynamics 
of pions, rather than heavy quarknoium itself, and will be separately 
discussed in Sec. IV.

In this talk, we review these recent developments in brief. In Section II, 
the NRQCD (and factorization formula) is introduced. In Section III, 
we discuss the color-octet mechanism for $J/\psi$ productions in various 
high energy  processes. In Section IV, the double peaks in $\Upsilon (3S) 
\rightarrow \Upsilon (1S) \pi\pi$ is discussed, and the summary is given 
in Section V. 

\section{NRQCD} 
\subsection{NRQCD Lagrangian}

A heavy quarkonium state $H$ is associated with  several different scales 
with the following hierarchical structure \cite{bbl95} : 
\begin{equation}
\Lambda_{QCD} \simeq M_Q v^2 << M_Q v << M_Q.
\end{equation}
In brief, $M_Q$ is the heavy quark mass scale that is a typical energy scale 
for decay and production of $H$, and PQCD in $\alpha_s (M_Q)$ becomes 
applicable. The  typical momentum scale of $Q$ inside $H$ is 
$p \sim M_Q v \sim 1 / {\rm size} $, and the typical kinetic energy scale
of $Q$ inside $H$ is $K.E. \sim M_Q v^2 \sim$ the level splittings. 
Finally, the typical energy scale for nonperturbative dynamics of 
light quarks and gluons is characterized by  $\Lambda_{\rm QCD}$.   
From the observed spectra of charmonium and upsilon families, one gets 
$v^2 \sim 0.3 $ for  the $\psi = (c \bar{c})$ system, and
$\sim 0.1 $ for the  $\Upsilon = (b \bar{b})$ system.
For sufficiently heavy $Q$, one has $v \sim \alpha_s (M_Q v) > 
\alpha_s (M_Q)$. Formally, the $v^2$ expansion is more important than 
$\alpha_s$ correction \cite{bbl95}.  

In order to obtain the NRQCD lagrangian, one integrates out the modes 
with momentum $> M_Q$ (no $Q/\bar{Q}$ creation/annihilation possible), 
and expand the resulting effective action in terms of heavy quark 
velocity. Thus NRQCD lagrangian posseses $SU(3)_c$ gauge symmetry, 
$P$ (parity) and $C$ (charge conjugation) symmetry, rotational symmetry, 
and approximate heavy quark spin symmetry. It is written in terms of 
heavy quark ($\psi$) and heavy antiquark fields ($\chi$), and 
their covariant derivatives :
\begin{equation}
{\cal L}_{\rm NRQCD} = \psi^{\dagger} \left( i D_0 + {D^2 \over 
2 M_Q} \right) \psi + \chi^{\dagger} \left( i D_0 - {D^2 \over 2 M_Q} 
\right) \chi + {\cal L}_{\rm light} + \delta {\cal L}
\end{equation}
Gluon exchange with $k \sim p \sim M_Q v$ is included in terms of static
potential, $A^0$, whereas soft gluons with $k \sim M_Q v^2$ are treated in
QCD multipole expansion.
 $\delta {\cal L}$ includes the correction terms of $O(M_Q v^4)$ and 
higher.
One can determine the coefficients of operators in NRQCD lagrangian 
by  calculating the same process in QCD and NRQCD, and expand both results 
in powers of $k/M_Q \sim v$, and match to the desired order in $v$ and 
$\alpha_s$.

In NRQCD, a physical heavy quarkonium state is represented as a 
superposition of various $(Q\bar{Q})_{1,8}$ and dynamical gluons :
$| J/\psi \rangle = O(1) | (c\bar{c})_{1} [ {^3S_1} ] \rangle +
O(v) | (c\bar{c})_{8} [ {^3S_1} ] g \rangle
+ O(v^2) | (c\bar{c})_{8} [ {^1S_0} ] \rangle
+ ...$. 
Relative importance of various Fock states in the above equation 
(and NRQCD matrix elements in factorization formulae discussed in the 
next subsection) are determined by velocity scaling laws that can be 
derived from field equation of motion  of NRQCD \cite{bbl95}.   

\subsection{NRQCD factorization}

Decays of a heavy quarkonium into light hadrons are described in terms of 
local 4-fermion operators in NRQCD lagrangian by optical theorem :
\begin{equation}
\Gamma (H \rightarrow LH)  =  2 {\rm Im}~\langle H | \delta 
{\cal L}_{\rm 4-fermion} | H \rangle
 =  \Sigma_n {2 {\rm Im} f_n(\Lambda) \over M_Q^{d_n - 4}}~\langle H | 
O_{n}(\Lambda) | H \rangle 
\end{equation}
This is a double expansion in $v^2$ (Non Pert.) and $\alpha_s (M_Q^2)$ 
(Pert.).


One can determine the NRQCD matrix elements either from
experimental data \cite{ko1} or lattice QCD \cite{skim}.
For $\chi_{c0}$ decay, the CMS prediction, (1), is modified (in the NRQCD 
formalism) into 
\begin{equation}
\Gamma ( \chi_{c0} \rightarrow {\rm LH} )  = 
{ \pi \alpha_s^2 (M_c)  \ M_c^5}~\langle O_{1} ({^3P_0}) 
\rangle  +  {N_f \pi \alpha_s^2 (M_c) \over 12 M_c^3}~
\langle  O_{8} ({^3S_1})  \rangle
\end{equation}
where
\begin{eqnarray}
\langle O_{1} \rangle = | \langle 0 | \chi^{\dagger} ( -{i\over 2} D 
\cdot \sigma ) \psi
\chi_{c0} \rangle |^2,
\\
\langle O_{8} \rangle =  \langle \chi_{c0} | \psi^{\dagger} \sigma T^a 
\chi \cdot \chi^{\dagger} \sigma T^a \psi | \chi_{c0} \rangle. 
\end{eqnarray}
Second term comes from the annihilation of the color-octet 
$(c\bar{c})_8$ ($S-$wave) in the physical $\chi_{c0}$ state.
IR div. in the first term at $O(\alpha_s^3)$ cacelled by the IR
div. in the color-octet matrix element in the second term. 
Similar cancellations were shown to  occur also in the $P-$wave charmonium 
production \cite{bbl92}.   

Inclusive cross section for a heavy quarkonium ($H$) production
\begin{equation}
\Sigma_{X} d\sigma (1+2 \rightarrow H(P) + X) 
= {1 \over 4 E_1 E_2 v_{12}}~{d^{3}P \over (2\pi)^3 2 E_P}~\Sigma_{mn}
C_{mn} \langle 0 | O_{mn}^{H} | 0 \rangle
\end{equation}
or,
\begin{equation}
d \sigma  = \Sigma_n d\hat{\sigma} ( Q\bar{Q}[n] + X ) \langle 0 | 
O^{H} (n) | 0 \rangle
\end{equation}
$C_{mn}$ takes into account the SD of order $1/M_c$ or less,
and therefore are calculable using PQCD in $\alpha_s (M_c)$.
The matrix element $\langle 0 | O_{mn}^{H} | 0 \rangle$ is the VEV of 
a  four-ferion operators of NRQCD, and is proportional to the 
inclusive transition probability of the perturbative $Q\bar{Q}[n]$ state 
into the physical heavy quarkonium state $H$.  
In terms of NRQCD operators, 
\begin{equation}
O_{mn}^H = \chi^{\dagger} {\cal K}_m^{' \dagger} \psi~{\cal P}_H ~
\psi^{\dagger} {\cal K}_n \chi,
\end{equation}
where ${\cal K}$'s are product of a spin matrix,  a color matrix and a 
polynomial in the covariant derivative, $D$, and the projection operator 
${\cal P}$ is defined as  
\begin{equation}
{\cal P}_H \equiv \Sigma_{S} |H(P=0,S\rangle \langle 
H(P=0),S|,
\end{equation}
where the sum is over soft hadron states $S$ with $E < \Lambda $.  

\section{Color-octet mechanism in $J/\psi$ productions} 
\subsection{$\psi^{'}$ anomaly at the Tevatron}

As discussed in the introduction, the color-octet gluon fragmentation 
$g \rightarrow (c\bar{c})_{8} [ {^3S_1} ] $ followed by 
$(c\bar{c})_{8} [ {^3S_1} ] \rightarrow J/\psi $ + (soft gluons) 
may explain the large excess of the $\psi^{'}$ at the Tevatron
\cite{fleming}. Qualitatively, one has 
\begin{eqnarray}
{\rm CSM} &  \sim \alpha_s^3 v^3 & ({\rm SD ~suppressed, LD ~enhanced}) 
\\
{\rm COM} &  \sim \alpha_s v^7 & ({\rm SD ~enhanced, LD~ suppressed})
\end{eqnarray}
Cho and Leibovich included the color-octet $^1S_0$ and $^3P_J$ contributions 
as well as the color-octet $^3S_1$  in terms of three NP parameters,
and obtained the following constraints on these parameters from the Tevatron 
data \cite{cho1} :
\begin{eqnarray*}
\langle 0 | O_{8}^{J/\psi} ({^1S_3}) | 0 \rangle  =  
(6.6 \pm 2.1) \times 10^{-3}~{\rm GeV}^3
\\
{\langle 0 | O_{8}^{J/\psi} ({^3P_J}) | 0 \rangle \over M_c^2}  +  
{\langle 0 | O_{8}^{J/\psi} ({^1S_0}) | 0 \rangle \over 3} 
 = 
(2.2 \pm 0.5) \times 10^{-2} ~{\rm GeV}^3
\end{eqnarray*}
%

It is very important to check the idea of the color-octet mechanism and NRQCD
factorization in other processes. 
In view of this, it is crucial to observe that the color-octet 
matrix elements $\langle 0 | O_{8}^{H} ({^{2S+1}L_J}) | 0 \rangle$ are
universal, i.e. process-independent. Therefore, one can determine these 
matrix elements from a (set of) process(es), and then apply to other 
processes, and test the NRQCD factorization
A lot of works have been done in this line. To name a few, 
hadroproduction of $h_c ({^1P_1})$ \cite{sridhar}, 
$B \rightarrow J/\psi + X$, 
$Z^0 \rightarrow J/\psi + X$ \cite{jungil}, 
$e^+ e^- \rightarrow J/\psi +X$ at CLEO \cite{swbaek},
to name only a few that were discussed at this workshop. 

\subsection{$B \rightarrow J/\psi +X$} 

The relevant effective Hamiltonian for this decay is 
\begin{equation}
H_{eff}  =  {G_E \over \sqrt{2}}~V_{cb} V_{cq}^{*}~ 
\left[  C_1 (\bar{c} c)_{V-A} 
(\bar{q} b)_{V-A} 
 +  C_2 (\bar{c} t^{a} c )_{V-A} (\bar{q} t^{a} b )_{V-A} \right],
\end{equation}
with $C_1 (m_b) \approx 0.13 $ and $C_{2} (m_b) \approx 2.21$ 
in the leading logarithmic approximation.
The CSM prediction to the lowest order in $\alpha_s$ is  
\begin{equation}
\Gamma ( B \rightarrow J/\psi + X )_{\rm csm}  
= {3 \langle 0 | O_{1}^{J/\psi} ({^3S_1}) 
| 0 \rangle \over 3 M_c^2}~C_1^2 
\left( 1 + { 8 M_c^2 \over M_b^2} \right) ~\hat{\Gamma}_0,
\end{equation}
with
$\hat{\Gamma}_0 \equiv |V_{cb}|^2 {G_F^2 \over 144 \pi}~M_b^3 M_c
~\left( 1 - { 4 M_c^2 \over M_b^2} \right)^2 $.
Using $\langle 0 | O_{1}^{J/\psi} ({^3S_1}) | 0 \rangle $ determined from 
$J/\psi \rightarrow l^+ l^-$, we get 
$B ( B \rightarrow J/\psi + X )_{\rm csm} = 0.23 \%$ 
compared with the most recent CLEO data
$B ( B \rightarrow J/\psi + X ) = ( 0.80 \pm 0.08) \% $.
Higher order corrections in $\alpha_s$ and relativistic corrections are 
not that important \cite{ko1}. Color-octet ${^3S_1}$ contribution
\begin{equation}
\Gamma ( b \rightarrow ( \bar{c} c )_{8} [{^3S_1}] + s \rightarrow J/\psi + 
X) = {\langle 0 | O_{8}^{J/\psi} ({^3S_1}) | 0 \rangle \over 2 M_c^2}
~C_2^2~\left( 1 + { 8 M_c^2 \over M_b^2} \right)
~\hat{\Gamma}_{0}
\end{equation}
is enhanced because of large Wilson coefficients \cite{ko1}, 
$C_2 \approx 17  C_1$. 
Thus, we get moderate increase in the branching ratio to $ 0.58 \%$.
Color-octet ${^1S_0}$ and ${^3P_J}$ also contribute at the same order
of $v^2$ expansion , dependent on 
$\langle 0 | O_{8}^{J/\psi} ({^1S_0}) | 0 \rangle $ and 
$\langle 0 | O_{8}^{J/\psi} ({^3P_J}) | 0 \rangle $. 
$J/\psi$ polarization in $B \rightarrow J/\psi + X$ \cite{nadeau}, which may
provide another useful constraint on the color-octet matrix elements. Similar 
analyses have been done for $B \rightarrow \chi_c +X$, and $B$ decays 
into $D-$wave charmonium state \cite{ko3}. Here the avaiable phase space is 
rather small so that the parton model description may be a poor 
approximation.

\subsection{$J/\psi$ Photoproduction}

The $J/\psi$ photoproduction ($\gamma + p \rightarrow J/\psi + X$) is 
described as a parton level subprocess $\gamma + g \rightarrow J/\psi + g$ 
in the PQCD and CSM \cite{berger}. 
This process is advocated as a nice probe of gluon distribution  function 
inside proton.  

PQCD corrections (in the CSM) to the leading order results have been done 
by M. Kr\"{a}mer \cite{kramer1}. 
The impacts of this radiative correction is that the scale 
dependence of $\alpha_s$ and structure functions is reduced. However,  
PQCD is out of control for $z > 0.8$ and $P_T^2 < 1$ GeV$^2$ at 
HERA energy ($\sqrt{s_{\gamma p}} = 100$ GeV. Therefore one needs   
cuts, $z < 0.8$ and $P_T^2 > 1$ GeV$^2$, when one employes the PQCD 
correction to the $J/\psi$ photoproduction.  

 Color-octet contributions were considered by various groups \cite{cacciari} 
\cite{ko2}. It was shown that $d\Gamma / dz$ spectrum in the high 
$z > 0.9$ region blows up, where $z$ is defined as $ E_{J/\psi} / 
E_{\gamma}$ in the proton rest frame.  This phenomenon was originally 
taken to be a signal that the determination of color-octet matrix elements 
from the Tevatron data on $J/\psi$ productions are inconsist with the 
$J/\psi$ photoproduction.

 However, this may not be the case because of breakdown of NRQCD 
near $z=1$ as discussed by Beneke et al. \cite{beneke}.  
In the lowest order in $v^2$, one ignores 
$\Delta M \equiv M_{H} - 2 M_Q$, {\it i.e.} set $M_{J/\psi} \approx 2 M_c$.
This is not valid near the phase space boundary, since one probes the 
dynamics in detail : $\Delta E < \Delta M$, especially, when the matrix 
element has mainly a support near the phase space boundary. 
This is the case for $J/\psi$ photoproduction at large $z (> 0.8 \sim 0.9)$. 
In such cases, predictions become sensitive to the $\Delta M$, or momentum 
carried away by light hadrons during the hadronization. One can summarize
this effect in terms of a universal shape function, as $b \rightarrow s 
\gamma$ in HQET. Shape function shifts the unphysical partonic boundary of 
phase space to the hadronic one that is physically more sensible. Therefore, 
the problem with $J/\psi$ photoproduction at $z>0.9$ is probably less 
serious.  

\section{$\pi\pi$ spectrum in $\Upsilon^{''} \rightarrow \Upsilon \pi\pi$}

The $M_{\pi\pi}$ spectra in  $\psi^{'} \rightarrow J/\psi \pi\pi$ and 
$\Upsilon^{'} \rightarrow \Upsilon \pi\pi$ can be understood in terms 
of QCD multipole expansion and the low energy theorem for pions, which 
dictates the following amplitude :   
\begin{equation}
{\cal M} = A ~\epsilon \cdot \epsilon^{'} ~\left[ q^2 + C M_{\pi}^2 \right],
\end{equation}
with $q^2 \equiv (p_1 + p_2 )^2 = M_{\pi\pi}^2$.  This amplitude predicts 
a peak at high $M_{\pi\pi}$ region in agreement with the data from ARGUS and 
CLEO.  However, the double humps in $\Upsilon (3S) (k,\epsilon) \rightarrow 
\Upsilon (1S) (k^{'}, \epsilon^{'}) \pi (p_1) \pi (p_2)$ could not be 
understood  with the above amplitude.  There are several proposals to this 
phenomenon, but none of them were successful. In Refs. ~\cite{chakrako}, the 
above amplitude was modified as 
\begin{equation}
{\cal M} = A ~\epsilon \cdot \epsilon^{'}~\left[ q^2 + B E_1 E_2 + 
C M_{\pi}^2 \right],
\end{equation}
and added the phase shift informations for $I=0$, $S$ and $D$ waves. Then, 
the authors of Refs.~\cite{chakrako} could fit the double humps, and 
predict various angular distributions. The predictions agree with the newest
CLEO data except the $\cos \theta_{\pi}^*$ distribution. It turns out that 
inclusion of  terms higher in pion momenta, such as 
$(E_1 + E_2 ) q^2$ and so on, improves the agreement with the data.
More systematic study in ChPT is in progress \cite{bai}.   

\section{Conclusion}

 NRQCD provides a theoretical framework in which one can study the 
perturbative (in $\alpha_s (M_Q)$) and nonperturbative (in $v^2$) 
aspects of heavy quarkonium physics.  In particular,  the role of the 
$| (Q\bar{Q})_8 g \rangle$ Fock state in the heavy quarkonium production 
and its decay can be rigorously formulated in the NRQCD.  
One can fit the $J/\psi$ and $\psi^{'}$ production rates at the Tevatron
via the color-octet mechanism.  
In order to check this idea at other processes, PQCD corrections are 
essential. This part has been calculated only recently by Petrelli 
{\it et al.}  \cite{petrelli}.  The complete phenomenological analysis 
including this new result has not been done yet, however.
Finally, Double hump in $M_{\pi\pi}$ spectrum can be explained in terms 
of an amplitude that satisfies the low energy theorem for pions, once 
the  $D-$wave  dipion amplitude is properly included.

\section*{Acknowledgements}

 I thank the organizing committe of this workshop, especially Manuel Drees 
and Kaoru Hagiwara, for their efforts to make this workshop successful and 
exciting. 
I am also grateful to Prof. H.S. Song, Jungil Lee, Seung Won Baek, 
Sun Myong Kim and S. Chakravarty for enjoyable collaborations on 
some materials presented in this talk.
This work was supported in part by KOSEF through CTP at Seoul National 
University, and by the Ministry of Education through the Basic Science 
Research Institute,  Contract No. BSRI-97-2418.

%
%

%
%

\end{document}